\documentclass[12pt]{iopart}

\usepackage{graphicx}

\begin{document}

\title[CMB Polarization in Theories of Gravitation with Massive Gravitons] {CMB Polarization in Theories of Gravitation with Massive Gravitons}

\author{Dennis Bessada\footnote[1]{dbessada@das.inpe.br}, Oswaldo D. Miranda\footnote[2]{oswaldo@das.inpe.br}} \address{INPE - Instituto Nacional de Pesquisas Espaciais -
Divis\~ao de Astrof\'isica, \\Av.dos Astronautas 1758, S\~ao Jos\'e dos
Campos, 12227-010 SP,
 Brazil\\}


\begin{abstract}

We study in this paper three different theories of gravitation
with massive gravitons - the modified Fierz-Pauli model, Massive
Gravity and the bimetric theory proposed by Visser - in linear
perturbation theory around a Minkowski and a flat
Friedmann-Robertson-Walker background. For the
transverse-traceless tensor perturbations we show that the three
theories give rise to the same dynamical equations, to the same
form of the tensor Sachs-Wolfe effect, and consequently to the
same form of the Boltzmann equations for the radiative transfer in
General Relativity.

We then analyze vector perturbations in these theories and show
that they do not give the same results as in the previous case. We
first show that vector perturbations in Massive Gravity present
the same form as found in General Relativity, whereas in the
modified Fierz-Pauli theory the vector gravitational-wave
polarization modes ($\Psi_{3}$ amplitudes in the Newman-Penrose
formalism) do not decay too fast as it happens in the former case.
Rather, we show that such $\Psi_{3}$ polarization modes give rise
to an unusual vector Sachs-Wolfe effect, leaving a signature in
the quadrupole form $Y_{2,\pm 1}(\theta,\varphi)$ on the Cosmic
Microwave Background Radiation polarization. We then derive the
details for the Thomson scattering of CMB photons for these
$\Psi_{3}$ modes, and then construct the correspondent Boltzmann
equations. Based upon these results we then qualitatively show
that $\Psi_{3}$-mode vector signatures - if they do exist - could
clearly be distinguished on the CMB polarization from the usual
$\Psi_4$ tensor modes.

We also estimate that the graviton mass limit for the vector modes
is $m=10^{-66}g\sim 10^{-29}cm^{-1}$, so that vector modes with
masses below this limit exhibit the same dynamical evolution as
the massless gravitons.

We argue at the end of this paper that CMB polarization
experiments can be decisive to test alternative theories of
gravitation by measuring CMB polarization in the $E$-mode.

\end{abstract}


\begin{center}
To appear in \emph{Classical and Quantum Gravity}
\end{center}

\maketitle

\section{\label{sec:intr}Introduction}

Among the most important basic predictions of the Theory of
General Relativity (GR), we know that only one of them has not
been yet directly tested: the existence of Gravitational Waves
(GWs). The GWs in the GR present two basic features: first, the
particle associated with the wave, the graviton, is massless;
second, they have only two polarization states. The first feature
comes exactly from the weak-field approximation in the space-time
metric, yielding a spin-two massless particle field equation; the
second feature comes from the gauge invariance of the field
equations under local coordinate transformations. These two
features are strictly related, since a massless theory allows
gauge transformations, and such transformations lead to two
polarization states.

However, a metric theory of gravitation, as we shall see, allows
up to six polarization modes in general \cite{lightman};
furthermore, it was shown in \cite{miranda} that spherical mass-resonant GW detectors might
detect up to six polarization modes. These two facts, one
theoretical and the other experimental, motivate us to investigate
more carefully a general metric theory of gravitation in order to
make it possible to predict further features that in the usual GR
might, in principle, be lacking.

An immediate generalization of GR may be constructed taking into
account some analogies with the basic ideas of Quantum Field
Theory (QFT). The simplest models in QFT usually involve free real
massless fields with global internal symmetries as, for example,
the electromagnetic field; then, to these models we can add
further elements like mass, non-abelian internal symmetries, and
so forth. Taking into account again the example of the
electromagnetic field, the next generalization of this model is
the introduction of a mass (via the Higgs mechanism), giving rise
to a massive spin-one particle. These two vector models, the
massless and the massive ones, are not theoretical conjectures;
they give rise to observable particles, the photon and the gauge
bosons $W^{\pm}$ and $Z^{0}$ respectively. Then, since in GR the
metric tensor field $g_{\alpha\beta}$ gives rise only to massless
particles, the next extension consists in the introduction of
massive degrees of freedom into the metric - the so-called
``\emph{massive gravitons}". There are some different effort
towards a theory of gravitation with massive gravitons; in this
paper we shall consider the approaches worked out in the
references \cite{Finn2002}, \cite{gruzinov1}, \cite{visser98},
\cite{rubakov2004} and \cite{dubovsky2004}.

The approach followed in references \cite{Finn2002} and
\cite{gruzinov1} is an improvement of an earlier model pushed
forward by M. Fierz and W. Pauli in the thirties \cite{fierz1939}.
It is constructed by adding a mass term to the linearized
Einstein-Hilbert action through the prescriptions of field theory,
that is, a quadratic term in the fields $h_{\alpha\beta}$
appearing in the weak-field limit. We shall call this model as
``modified Fierz-Pauli" henceforth. In this model, GWs have
\emph{six} polarization modes \cite{boulware}, which may introduce
interesting features in cosmology as we are going to show in this
paper.

The second approach, devised by Matt Visser, \cite{visser98} is
based upon a bimetric theory, first pushed forward by the work of
Rosen \cite{rosen1973}. In this model the mass term is introduced
by a quadratic term depending not only upon a dynamical metric as
in GR, but upon a nondynamical background metric as well. De Paula
\textit{et. al.} \cite{wayne2004} showed that this theory leads to
exactly the same dynamical equations as in the modified
Fierz-Pauli model in the weak-field approximation, so that they
are absolutely equivalent in this limit. They also have shown that
GWs in this model also possess six polarization modes as expected
\cite{boulware}.

In the theory of \emph{Massive Gravity}, as introduced in
\cite{rubakov2004}, the Lorentz-invariance of the mass lagrangian
is broken in the following way: the quadratic term in the metric
perturbations is split into components, giving rise to five
possible combinations, with each combination having a different
coefficient; then, these five different coefficients are
interpreted as five \emph{mass parameters}, each of them being
proportional to a common scale $m$. It was also shown in
\cite{rubakov2004} that the Fierz-Pauli model is promptly
recovered through a suitable choice of the mass parameters. In
\cite{dubovsky2004}, the masses for the gravitons are introduced
through a very clever analogy with the Higgs mechanism in QFT: the
Lorentz invariance is spontaneously broken by a convenient choice
of the ``vacuum"$~$for the Goldstone fields, which leads to the
model devised in \cite{rubakov2004}.

In all the models introduced above the van Dam-Veltmann-Zakharov (vDVZ)
discontinuity \cite{Veltman1970}, \cite{zakharov} is absent \cite{gruzinov1}, \cite{visser98}, \cite{rubakov2004}, \cite{dubovsky2004}. In particular, it can be shown that Massive Gravity is absolutely free of ghosts
and classical instabilities provided the mass parameters obey some
constraints \cite{dubovsky2004}.

In addition to the efforts towards a direct observation of GWs
(see, \textit{e.g.}, ref. \cite{ligo}), we might consider, as an
excellent alternative for the time being (since the GW detectors
have not yet reached the proper sensitivity to make such direct
observations), an indirect approach, using the recent work
performed on the theory of Cosmic Microwave Background Radiation
(CMB) anisotropies and polarization. The CMB spectrum might hide
some ``tracks" of GWs, or \emph{signatures}, left by the
interaction of the primordial cosmological gravitons with the CMB
photons. The next generation of CMB satellites after Planck is
expected to measure the CMB polarization to a better degree of
accuracy, and it can hopefully shed a light on the investigations
of primordial GWs through the measurements of the so-called E and
B-polarization modes \cite{bmode}. This means that CMB
polarization measurements might not only be decisive to detect
GWs, but it might also shed a light on the nature of the gravity
itself; in other words, CMB measurements could be decisive to test
alternative theories of gravitation. This is exactly our goal in
the present work: the analysis of CMB polarization induced by GWs
in a theory of gravitation with massive gravitons.

To this end, the present paper is organized as follows: in Section
\ref{sec:pol} we review the classification of plane GWs in the
Newman-Penrose formalism for an arbitrary theory of gravitation,
deducing their six possible polarization modes. In Section
\ref{sec:massgrav} we review the three different approaches to
include massive gravitons into GR, and also discuss cosmological
perturbations in these theories. Then, in Section \ref{sec:cmbpol}
we review the basics of radiative transfer in the presence of weak
gravitational fields, laying the ground for discussing the
Sachs-Wolfe effect in a theory of gravitation with massive
gravitons in Section \ref{sec:sachswolfe}, and the effect of GW
vector longitudinal modes on the Thomson scattering in Section
\ref{sec:thomscat}. In Section \ref{sec:boltzeq} we get together
all the results, obtaining the related radiation transfer
equations (Boltzmann equations) for each polarization mode. At the
end of this paper we discuss the obtained results and make the
correspondent conclusions.

\section{\label{sec:pol}Polarization States for an arbitrary Metric
 Theory of Gravitation}

The polarization states for a GW in an arbitrary metric theory of
 gravitation are given by the independent modes of the Riemann tensor. In
 order to compute its components in an Lorentz-invariant scheme, it is
 convenient to introduce, following the pioneering work of Newman and Penrose,
 \cite{penrose}, the quasiorthonormal complex-null basis
 $(k,l,m,\bar{m})$, where $k$ and $l$ are real null-vectors and $m$ and $\bar{m}$ are a
 pair of complex numbers, satisfying the following orthogonality
 relations:
\begin{equation}\label{npbasis}
k\cdot l=0, ~~~ m\cdot \bar{m}=-1,~~~ k\cdot \bar{m}=k\cdot m
=l\cdot \bar{m}=l\cdot m=0.
\end{equation}
We follow \cite{lightman} and choose the following set of null
vectors,
\begin{eqnarray}
k&=&-\frac{1}{\sqrt{2}}(1,0,0,1),~~~l=-\frac{1}{\sqrt{2}}(1,0,0,-1),\label{nullvectorsk}\\
m&=&-\frac{1}{\sqrt{2}}(0,1,i,0),~~~\bar{m}=-\frac{1}{\sqrt{2}}(0,1,-i,0),\label{nullvectorsmbar}
\end{eqnarray}
all satisfying (\ref{npbasis}). With the basis given by
(\ref{nullvectorsk}) - (\ref{nullvectorsmbar}) we can split the
Riemann tensor into its irreducible parts, namely, the Weyl
tensor, whose ten independent components are given by five complex
scalars $(\Psi_{0},\Psi_{1},\Psi_{2},\Psi_{3},\Psi_{4})$, the
Ricci tensor, whose nine independent components are given by the
scalars $\Phi_{00}$, $\Phi_{01}$, $\Phi_{02}$, $\Phi_{10}$,
$\Phi_{20}$, $\Phi_{11}$, $\Phi_{12}$, $\Phi_{21}$, $\Phi_{22}$,
and the Ricci scalar $\Lambda$. Throughout this section, we
consider only plane GW propagating in the $\mathbf{\hat{z}}$
direction, whose time dependence is given by $\cos \omega t$, and
we use for the flat Minkowski metric
$\eta_{\alpha\beta}=$diag$\{+,-,-,-\}$.

In this context, we may prove that the differential and algebraic
properties of the Riemann tensor reduce the number of independent
components to six \cite{lightman}, given by
\\
\\
\textit{i})
\textit{The Weyl tensor}:
\begin{eqnarray}
\Psi_{0}&=&\Psi_{1}=0, \label{psi01}\\
\Psi_{2}&=&-\frac{1}{6}R_{lklk},\label{psi2}\\
\Psi_{3}&=&-\frac{1}{2}R_{lkl\bar{m}}, \label{psi3}\\
\Psi_{4}&=&-R_{l\bar{m}l\bar{m}}\label{psi4};
\end{eqnarray}

\textit{ii}) \textit{The Ricci tensor}:

\begin{eqnarray}
\Phi_{00}&=&\Phi_{01}=\Phi_{10}=\Phi_{02}=\Phi_{20}=0,\label{phi00}\\
\Phi_{22}&=&-R_{lml\bar{m}},\label{phi22} \\
\Phi_{11}&=&\frac{3}{2}\Psi_{2},\label{phi11} \\
\Phi_{12}&=&\bar{\Phi}_{21}=\bar{\Psi}_{3};\label{phi12}
\end{eqnarray}
\\
\textit{ii}) \textit{The Ricci scalar}:
\begin{equation}\label{lambda}
\Lambda=-\frac{1}{2}\Psi_{2}.
\end{equation}

We can reduce, therefore, the number of independent components of
the Riemann tensor to the set
\begin{equation}\label{npamplitudes}
\{\Psi_{2},\Psi_{3},\bar{\Psi}_{3},\Psi_{4},\bar{\Psi}_{4},\Phi_{22}\}.
\end{equation}

Henceforth, we call (\ref{npamplitudes}) \textit{Newman-Penrose}
\textit{(NP)} \textit{amplitudes}. They play the role of definite
helicity states $s=(0,\pm 1,\pm 2)$ under rotations around the z
axis in a nearly Lorentz coordinate frame. In particular, the two
real NP amplitudes $(\Psi_{2},\Phi_{22})$ correspond to the state
$s=0$ (which defines the scalar modes), whereas the complex NP
amplitudes $(\Psi_{3},\bar{\Psi}_{3})$ correspond to $s=\pm1$
(vector modes), and $(\Psi_{4},\bar{\Psi}_{4})$ to $s=\pm2$
(tensor modes). These polarization modes can be represented on the
$x-y$, $y-z$ or $x-z$ plane as can be seen through the effects of
a GW on a ring of dust particles (see reference \cite{will1} for
further details). Now, following \cite{lightman}, it is very
useful to introduce the ``driving-force matrix" $S$,
\begin{equation}\label{drivingforcematrix}
S_{ij}(t):=R_{i0j0}(u),
\end{equation}
where $t$ is the proper time and $u=t-z/c$ represents a null
``retarded time" as measured by an ideal detector in the
coordinate system $\{t,x^{i}\}$. From (\ref{drivingforcematrix})
we define a basis for the GW polarizations as follows: first, we
represent the NP amplitudes as
\begin{eqnarray}
p_{1}(\hat{z},t)&=&\Psi_{2}(u),\label{p1}~~~
p_{2}(\hat{z},t)=\textrm{Re} ~\Psi_{3}(u),~~~
p_{3}(\hat{z},t)=\textrm{Im} ~\Psi_{3}(u),\\
p_{4}(\hat{z},t)&=&\textrm{Re} ~\Psi_{4}(u),~~~
p_{5}(\hat{z},t)=\textrm{Im} ~\Psi_{4}(u),~~~
p_{6}(\hat{z},t)=\Phi_{22}(u);\label{p6}
\end{eqnarray}
(from now on, whenever the \emph{polarization index} $r$ appears,
it will always indicate a given NP amplitude according to the
sequence given in (\ref{p1}-\ref{p6}), so that $r=1$ stands for
$\Psi_{2}$, and so forth). Now, writing the NP amplitudes
(\ref{psi2}), (\ref{psi3}), (\ref{psi4}) and (\ref{phi22}) in
Cartesian coordinates (recall that the NP basis given by
(\ref{nullvectorsk}) - (\ref{nullvectorsmbar}) can be written in
terms of the coordinates $\{t,x^{i}\}$), we get the following
result
\begin{eqnarray}\label{drivingforcematrixcomp}
S&=&\left(
\begin{array}{ccc}
-\frac{1}{2}(p_{4}+p_{6}) & \frac{1}{2}p_{5} & -2p_{2} \\
\frac{1}{2}p_{5} & \frac{1}{2}(p_{4}-p_{6}) & 2p_{3} \\
-2p_{2} & 2p_{3} & -6p_{1} \\
\end{array}
\right)\\
&=&\sum_{r=1}^{6}p_{r}\left(\hat{z},t\right)E_{r}\left(\hat{z}\right),
\end{eqnarray}
where $E_{r}\left(\hat{z}\right)$ are the \textsl{basis polarization
matrices}, given by
\begin{eqnarray}\label{polbasis}
E_{1}&=&-6\left(
\begin{array}{ccc}
0 & 0 & 0 \\
0 & 0 & 0 \\
0 & 0 & 1 \\
\end{array}
\right), ~~ E_{2}=-2\left(
\begin{array}{ccc}
0 & 0 & 1 \\
0 & 0 & 0 \\
1 & 0 & 0 \\
\end{array}
\right)
\nonumber \\
E_{3}&=&2\left(
\begin{array}{ccc}
0 & 0 & 0 \\
0 & 0 & 1 \\
0 & 1 & 0 \\
\end{array}
\right), ~~ E_{4}=-\frac{1}{2}\left(
\begin{array}{crc}
1 & 0 & 0 \\
0 & -1 & 0 \\
0 & 0 & 0 \\
\end{array}
\right), \nonumber \\ E_{5}&=&\frac{1}{2}\left(
\begin{array}{ccc}
0 & 1 & 0 \\
1 & 0 & 0 \\
0 & 0 & 0 \\
\end{array}
\right), ~~~ E_{6}=-\frac{1}{2}\left(
\begin{array}{ccc}
1 & 0 & 0 \\
0 & 1 & 0 \\
0 & 0 & 0 \\
\end{array}
\right).
\end{eqnarray}

Therefore, the polarization of a GW in an arbitrary metric theory
of gravity can be fully described by the basis polarization
matrices $E_{r}\left(\hat{z}\right)$. However, due to the
tensorial character of the space-time metric it is convenient to
cast the polarization basis (\ref{polbasis}) into a tensor; hence,
along with its spatial components, given by
$\left(E_{r}\right)_{ij}\left(\hat{z}\right)$, there are the $00$
and $0i$ components, which are zero by the very definition of the
``full driving-force matrix"$~S$ (\ref{drivingforcematrix}), so
that
\begin{equation}\label{drivingforcematrix1}
S_{00}(t)=R_{0000}(u)=0, ~~~~
S_{0i}(t)=R_{0i00}(u)=0.
\end{equation}
Hence, the \emph{polarization tensor} assumes the form

\begin{eqnarray}\label{polariztensor}
\varepsilon^{1}&=&\left(
\begin{array}{cccc}
0 & 0 & 0 & 0 \\
0 & 0 & 0 & 0 \\
0 & 0 & 0 & 0 \\
0 & 0 & 0 & 1 \\
\end{array}
\right), ~~ \varepsilon^{2}=\left(
\begin{array}{cccc}
0 & 0 & 0 & 0 \\
0 & 0 & 0 & 1 \\
0 & 0 & 0 & 0 \\
0 & 1 & 0 & 0 \\
\end{array}
\right) \nonumber \\
\varepsilon^{3}&=&\left(
\begin{array}{cccc}
0 & 0 & 0 & 0 \\
0 & 0 & 0 & 0 \\
0 & 0 & 0 & 1 \\
0 & 0 & 1 & 0 \\
\end{array}
\right), ~~ \varepsilon^{4}=\left(
\begin{array}{ccrc}
0 & 0 & 0 & 0 \\
0 & 1 & 0 & 0 \\
0 & 0 & -1 & 0 \\
0 & 0 & 0 & 0 \\
\end{array}
\right), \nonumber \\
\varepsilon^{5}&=&\left(
\begin{array}{cccc}
0 & 0 & 0 & 0 \\
0 & 0 & 1 & 0 \\
0 & 1 & 0 & 0 \\
0 & 0 & 0 & 0 \\
\end{array}
\right), ~~ \varepsilon^{6}=\left(
\begin{array}{cccc}
0 & 0 & 0 & 0 \\
0 & 1 & 0 & 0 \\
0 & 0 & 1 & 0 \\
0 & 0 & 0 & 0 \\
\end{array}
\right).
\end{eqnarray}

\section{\label{sec:massgrav}The Theory of Gravitation with Massive Gravitons}

\subsection{\label{sec:modfp} The modified Fierz-Pauli model}

Let us now analyze how do GWs arise in the case of the Fierz-Pauli
modified model. In this case, the graviton mass lagrangian appears
as a quadratic term in the perturbation of the metric tensor
$h_{\alpha\beta}$ in the weak-field limit, so that its action is
given by \cite{Finn2002}, \cite{gruzinov1}:
\begin{eqnarray}\label{action1}
S&=&\frac{M_{Pl}^{2}}{8}\int
d^{4}x\left[h_{\alpha\beta,\gamma}h^{\alpha\beta,\gamma}
-2{h_{\alpha\beta}}^{,\beta}{h^{\alpha\gamma}}_{,\gamma}+
2{h_{\alpha\beta}}^{,\beta}{h^{,\alpha}}-h^{,\alpha}h_{,\alpha}\right.
\nonumber \\
&-& \left. 4M_{Pl}^{-2}h_{\alpha\beta}T^{\alpha\beta} - m^2\left(h_{\alpha\beta}h^{\alpha\beta}-\frac{1}{2}
h^2\right)\right],
\end{eqnarray}
where $h$ is given by
\begin{equation}\label{htrace}
h=\eta_{\alpha\beta}h^{\alpha\beta},
\end{equation}
and $M_{Pl}$ is the Planck mass. If instead of the contribution
$m^2h^2/2$ to the last term on the right-hand side of
(\ref{action1}) one had $m^2h^2$, this model would correspond to
the original Fierz-Pauli action, which is plagued by the vDVZ
discontinuity \cite{gruzinov1}.

The Einstein equations associated with the action (\ref{action1})
are given by
\begin{eqnarray}\label{eqmotion}
\partial^{\mu}\partial_{\mu}
h_{\alpha\beta}&-&{{h_{\alpha}}^{\gamma}}_{,\gamma\beta}-{{h_{\beta}}^{\gamma}}_{,\gamma\alpha}
+h_{,\alpha\beta}+\eta_{\alpha\beta}{h^{\gamma\delta}}_{,\gamma\delta}
\nonumber \\ &-&\eta_{\alpha\beta}
\partial^{\mu}\partial_{\mu}
h+
 m^2\left(h_{\alpha\beta}-\frac{1}{2}\eta_{\alpha\beta}h\right)=- 2M_{Pl}^{-2}T_{\alpha\beta};
\end{eqnarray}
then, imposing the conservation of the stress energy-momentum
tensor, $\nabla_{\alpha} T^{\alpha\beta}=0$, we get the following
constraint to the field $h_{\alpha\beta}$ on a Minkowski
background,
\begin{equation}\label{hbarconstraint}
\partial_{\alpha}\bar{h}^{\alpha\beta}=0,
\end{equation}
where we have defined
\begin{equation}\label{hbardef}
\bar{h}_{\alpha\beta}=h_{\alpha\beta}-\frac{1}{2}\eta_{\alpha\beta}h.
\end{equation}

The equation (\ref{hbarconstraint}) is exactly the same found in
GR, but in the present case it emerges as a constraint from the
conservation of the stress energy-momentum tensor rather than a
gauge choice, as in GR. This constraint eliminates four degrees of
freedom out of the ten independent components of the space-time
metric, leaving then only six independent modes. Since these modes
correspond exactly to the polarization states of the GW, we may
readily associate the components of ${\bar{h}}_{\alpha\beta}$ with
the correspondent ones of (\ref{polariztensor}), so that the only
nonzero contributions are the spatial components $\bar{h}_{ij}$.

Using the arguments above and plugging equation
(\ref{hbarconstraint}) into (\ref{eqmotion}), we obtain, in the
absence of sources,
\begin{equation}\label{waveq}
\left(\partial^{\mu}\partial_{\mu} + m^2\right){\bar{h}}_{ij}=0,
\end{equation}
which is clearly a Klein-Gordon equation for a wave propagating in
the direction $\hat{\mathbf{k}}=\mathrm{\hat{z}}$. For the sake of
simplicity we henceforth drop the bar over the tensor on the
left-hand side of (\ref{hbardef}) and simply write it as $h_{ij}$.

Due to the oscillatory character of equation (\ref{waveq}) we may expand the tensor field $h_{ij}$ into the Fourier modes as follows,
\begin{eqnarray}\label{fourierexph}
h_{ij}\left(x\right)=\int^{\infty}_{-\infty}
\frac{d^{3}k}{(2\pi)^{3/2}}{\tilde{h}}_{ij}\left(k\right)e^{-ikx},
\end{eqnarray}
which enables us to write down the following decomposition in terms of the polarization tensor (\ref{polariztensor}):
\begin{eqnarray}\label{expanh}
{\tilde{h}}_{ij}\left(k\right)&=&\sum_{r=1}^{6}\varepsilon^{r}_{ij}(k)\tilde{h}^{r}(k).
\end{eqnarray}
In particular, for the transverse-traceless (TT) component of
the tensor perturbation to the metric $h_{ij}$ (corresponding to
the $\Psi_{4}$ mode with $r=4,5$), we write
\begin{equation}\label{httff}
{\tilde{h}}^{\perp}_{ij}=\varepsilon^{4}_{ij}(k)\tilde{h}^{4}(k)+\varepsilon^{5}_{ij}(k)\tilde{h}^{5}(k),
\end{equation}
so that Fourier transforming (\ref{httff}) to get ${h}^{\perp}_{ij}$ in the configuration space, we see that this tensor satisfies (\ref{waveq}), that is
\begin{equation}\label{waveqtt}
\left(\partial^{\mu}\partial_{\mu} + m^2\right){h}^{\perp}_{ij}=0,
\end{equation}
which reduces to the GW equation for GR in the limit $m=0$. The
tensor ${h}^{\perp}_{ij}$ encompasses then both transverse
polarization modes ``$+$" and ``$\times$" characteristic of GR.

We write the extension of definition (\ref{httff}) to the $\Psi_3$ modes (associated with $r=2,3$)
as
\begin{equation}\label{hvec}
{\tilde{h}}^{\parallel}_{ij}=\varepsilon^{2}_{ij}(k)\tilde{h}^{2}(k)+\varepsilon^{3}_{ij}(k)\tilde{h}^{3}(k),
\end{equation}
which corresponds to a longitudinal polarization state. As we
shall see in Section \ref{sec:sachswolfe}, these modes can also
induce an unusual angular pattern for the CMB photons and, in
consequence, induce a different signature in its polarization
pattern.

In this paper we do not consider the GW scalar polarization modes
$\Psi_2$ and $\Phi_{22}$, since they couple to the $\delta g_{00}$
scalar component in the metric perturbation on cosmological
scales, and then do not produce ``handedness" to excite the CMB
B-polarization mode.

\subsection{\label{sec:massivegravity} The bimetric model}

The bimetric model proposed by Visser in \cite{visser98} combines
into a single theory a dynamical space-time metric with a
nondynamical metric which allows the introduction of a mass for
the graviton even in the strong-field limit. However, as was shown
by De Paula \emph{et. al.} \cite{wayne2004}, in the weak-field
limit the bimetric model is absolutely equivalent to the modified
Fierz-Pauli model as discussed above. Due to this property we
henceforth consider the modified Fierz-Pauli solely.

\subsection{\label{sec:massivegravity} Massive Gravity}

In the case of Massive Gravity, as we have pointed out in the
Introduction, the underlying ideas are a bit different. Let us now
write down the space-time metric in the weak-field limit as
\begin{equation}\label{linearapprox}
g_{\alpha\beta}=\eta_{\alpha\beta}+\delta g_{\alpha\beta},
\end{equation}
where $\delta g_{\alpha\beta}$ plays the role of a perturbation to the Minkowski background metric, and assumes the form \cite{rubakov2004},
\begin{eqnarray}
\delta g_{00}=2\varphi, ~~~ \delta g_{0i}=S_{i}-\partial_{i}B,\nonumber
\end{eqnarray}
\begin{equation}\label{pertmetricmassgrav}
\delta
g_{ij}=-\chi_{ij}-\partial_{i}F_{j}-\partial_{j}F_{i}+2\left(\psi\delta_{ij}-\partial_{i}\partial_{j}E\right),
\end{equation}
where $\varphi,\psi,B,E$ are scalar fields, $F_{i}$ and $S_{i}$
are vector fields, and $\chi_{ij}$ is a tensor field. The vector and tensor fields of (\ref{pertmetricmassgrav}) satisfy the well known constraints
\begin{equation}\label{pertmetricconstr}
{\chi_{ij}}^{,~j}=0,~~~
{\chi^{i}}_{i}=0,~~~{F^{i}}_{,i}={S^{i}}_{,i}=0
\end{equation}
necessary to match the number of independent fields to the ten
independent components of the metric $\delta g_{\alpha\beta}$.
Now, we construct the mass contribution to the action of the
theory as a quadratic term in the tensor fields as well, but
breaking the Lorenz invariance as suggested in \cite{rubakov2004}
\begin{equation}\label{masslagr1}
{\cal{L}}_{m}=\frac{M_{Pl}^{2}}{2}\left[m_{0}^{2}\delta g_{00}^{2}
+ 2m_{1}^{2}\delta g_{0i}^{2}-m_{2}^{2}\delta g_{ij}^{2}
+m_{3}^{2}\delta g_{ii}\delta g_{jj}-2m_{4}^{2}\delta g_{00}\delta
g_{ii}\right].
\end{equation}
Now, plugging the decomposition (\ref{pertmetricmassgrav}) into
the mass lagrangian (\ref{masslagr1}) and in the usual
Einstein-Hilbert lagrangian of GR, and adding up these terms, we
get the following equation for the tensor field $\chi_{ij}$
\cite{rubakov2004}
\begin{equation}\label{waveqchi}
\left(\partial^{\mu}\partial_{\mu} + m_{2}^2\right){\chi}_{ij}=0,
\end{equation}
which is exactly the same equation for the TT metric contribution of the modified Fierz-Pauli model given by (\ref{waveqtt}), representing the ``genuine" GW polarization modes for Massive Gravity. As we have argued in Section \ref{sec:modfp}, the $\Psi_3$ content of the modified Fierz-Pauli model plays the role of extra GW longitudinal polarization modes; however, in Massive Gravity, the vector fields $F_{i}$ and
$S_{i}$ evolve as massive spin-one particles with transverse polarization, and have nothing to do with extra GW polarization states.

Besides this original approach devised by V. Rubakov
\cite{rubakov2004}, there is another way of breaking the Lorentz
symmetry of the massive lagrangian which resembles the Higgs
mechanism in QFT. In this case, we introduce a set of four
Goldstone fields $\phi^{0}(x)$ and $\phi^{i}(x)$, $i=1,2,3$, and a
scalar, a vector and a tensor field constructed as follows
\cite{dubovsky2004}, \cite{dubovsky2005a}:
\begin{equation}\label{xfield}
X=\Lambda^{-4}g^{\alpha\beta}\partial_{\alpha}\phi^{0}\partial_{\beta}\phi^{0},
\end{equation}
\begin{equation}\label{vfield}
V^{i}=\Lambda^{-4}g^{\alpha\beta}\partial_{\alpha}\phi^{0}\partial_{\beta}\phi^{i},
\end{equation}
\begin{equation}\label{wfield}
W^{ij}=\Lambda^{-4}g^{\alpha\beta}\partial_{\alpha}\phi^{i}\partial_{\beta}\phi^{j}-X^{-1}V^{i}V^{j},
\end{equation}
where $\Lambda$ is the energy scale of the theory. With such
elements we introduce an arbitrary function
$F=F\left(X,V^{i},W^{ij}\right)$, so that the action for Massive
Gravity reads
\begin{equation}\label{massgravaction}
S=\int d^{4}x
\sqrt{-g}\left[-M_{Pl}^{2}R+\Lambda^{4}F(X,V^{i},W^{ij})+{\cal{L}}_{matter}\right],
\end{equation}
where the first term on the right-hand side represents the usual
Einstein-Hilbert action, and ${\cal{L}}_{matter}$ is the
lagrangian for ordinary matter minimally coupled to the metric. We
can show  that in the linear perturbation regime given by
(\ref{linearapprox}), after setting the Goldstone fields to their
``vacuum" values,
\begin{equation}\label{goldsvac}
g_{\alpha\beta}=\eta_{\alpha\beta}, ~~~ \phi^{0}=\Lambda^{2}t,
~~~\phi^{i}=\Lambda^{2}x^{i},
\end{equation}
and substituting these values and the decomposition
(\ref{pertmetricmassgrav}) into (\ref{massgravaction}) we get
exactly the same mass lagrangian as (\ref{masslagr1}), with the
mass parameters now being related to the functions $F$ and their
derivatives \cite{dubovsky2004}, \cite{dubovsky2005a}.

\subsection{\label{sec:cosmopert}Cosmological Perturbations for Massive Gravitons}

Once we have discussed the key features of metric perturbations
around a flat Minkowski background, let us now address the same
question in a flat Friedmann-Robertson-Walker (FRW) background. We
start with GR in which, in the standard theory of cosmological
perturbations \cite{brandenberger1992}, the metric $\delta
g_{\alpha\beta}$ is decomposed exactly in the same way as we did
in (\ref{pertmetricmassgrav}) but, in this case, we multiply all
these components by the square of the scale factor of the universe
$a(\eta)$ ($\eta$ is the conformal time), that is
\begin{equation}\label{cosmopertmetricmassgrav}
\delta g_{\alpha\beta}=a(\eta)^{2}\left(
\begin{array}{cc}
2\varphi & S_{i}-\partial_{i}B  \\
S_{i}-\partial_{i}B & -\chi_{ij}-\partial_{i}F_{j}-\partial_{j}F_{i}+2\psi\delta_{ij}-2\partial_{i}\partial_{j}E \\
\end{array}
\right).
\end{equation}
The constraints for the cosmological case are the same as
(\ref{pertmetricconstr}). Now, since we are interested in the CMB
polarization induced by GWs, we focus only on the TT part of the
metric perturbation; in this case, the Einstein equation for the
tensor field is given by (the prime $'$ indicates a derivative
with respect to conformal time)
\begin{equation}
{{h''}}_{ij}-\nabla^2 {{h}}_{ij}+2\mathcal{H}{{h}}_{ij}'
=0,
\label{tensoreinsteinperttensoriaisnull}
\end{equation}
which simply describes transverse GW travelling on an expanding
background.

In the case of the modified Fierz-Pauli model the same metric
decomposition cannot be performed due to the extra polarization
modes; we instead introduce
\begin{equation}
\label{pertmetmodfp}
\delta g_{\alpha\beta}=a(\eta)^{2}\left(
\begin{array}{cc}
2\phi & X_{i}-Q_{,i}  \\
X_{i}-Q_{,i} & - h_{ij} \\
\end{array}
\right),
\end{equation}
where $\phi$ and $Q$ are scalar fields, $X_i$ is a divergenceless
vector field, and $h_{ij}$ is the cosmological version of the
tensor given by the solution to equation (\ref{waveq}), carrying
the correspondent six polarization modes spanned in the NP
formalism. The two scalar fields, plus the two components of the
transverse vector field and the six modes of the tensor field give
exactly the required ten degrees of freedom. The mass lagrangian
for this model can be constructed analogously as in
(\ref{action1}), that is, it appears as a quadratic term in the
metric (\ref{pertmetmodfp}). The full action is then obtained by
adding up this contribution to the usual Einstein-Hilbert one, and
the Einstein equations can be derived using the standard tools.
Before doing that, it is convenient to decompose the tensor
perturbation $h_{ij}$ into its TT and longitudinal parts in the
Fourier space. In (\ref{fourierexph}) the whole time-dependence of
$h_{ij}$ is contained in the exponential since it is a solution to
a wave equation of the form (\ref{waveq}); now, such
time-dependence changes because of the extra temporal function
$a(\eta)$ appearing in (\ref{pertmetmodfp}), which introduces a
damping in the oscillation. Therefore, we Fourier-expand the
massive tensor perturbation $h_{ij}$ as
\begin{eqnarray}\label{cosmofourierexph}
h_{ij}\left(x\right)&=&\int^{\infty}_{-\infty}
\frac{d^{3}\mathbf{k}}{(2\pi)^{3/2}}{\tilde{h}}_{ij}\left(\eta,\mathbf{k}\right)e^{-i\mathbf{k}\cdot
\mathbf{r}},
\end{eqnarray}
where
\begin{eqnarray}\label{cosmofourierexph1}
\tilde{h}_{ij}\left(\eta,\mathbf{k}\right)&=&\sum^{6}_{r=1}\varepsilon^{r}_{ij}(\mathbf{k})\tilde{h}^{r}
\left(\eta,\mathbf{k}\right),
\end{eqnarray}
so that the TT and longitudinal components of
${\tilde{h}}_{ij}$ can be written in the same foot as
(\ref{httff}) and (\ref{hvec}), that is
\begin{eqnarray}\label{httffcosmo}
{\tilde{h}}^{\perp}_{ij}\left(\eta,\mathbf{k}\right)&=&\varepsilon^{4}_{ij}(\mathbf{k})\tilde{h}^{4}
\left(\eta,\mathbf{k}\right)+
\varepsilon^{5}_{ij}(\mathbf{k})\tilde{h}^{5}\left(\eta,\mathbf{k}\right),\\
{\tilde{h}}^{\parallel}_{ij}\left(\eta,\mathbf{k}\right)&=&\varepsilon^{2}_{ij}(k)
\tilde{h}^{2}\left(\eta,\mathbf{k}\right)
+\varepsilon^{3}_{ij}\left(\mathbf{k}\right)\tilde{h}^{3}\left(\eta,\mathbf{k}\right);
\end{eqnarray}
now, extracting the Einstein equations from the action for the cosmological Fierz-Pauli model as we have sketched above, we see that both fields $h^{\perp}_{ij}$ and $h^{\parallel}_{ij}$ satisfy the same dynamical equations
\begin{eqnarray}\label{tttensorpertmfp}
{h^{\perp''}}_{ij}-\nabla^2 {h^{\perp}}_{ij}+2{\cal{H}}h^{\perp'}_{ij}+a^{2}m^{2}h^{\perp}_{ij}&=&0,\\
{h^{\parallel''}}_{ij}-\nabla^2 {h^{\parallel}}_{ij} +2{\cal{H}}{h^{\parallel}_{ij}}'+a^{2}m^{2}h^{\parallel}_{ij}&=&0\label{vectensorpertmfp}.
\end{eqnarray}

In the case of Massive Gravity, the cosmological perturbations to
the metric are given by (\ref{cosmopertmetricmassgrav}), together
the following set of perturbations to the Goldstone fields in the
unitary gauge (\ref{goldsvac}),
\begin{equation}
\label{transfgoldstone}
\tilde{\phi^{0}}=\phi^{0}+\Lambda^2\lambda^{0}, ~~~~~
\tilde{\phi}^{i}=\phi^{i}+\Lambda^2\left(\lambda^{i}+\lambda^{,i}\right),
\end{equation}
where $\lambda^{0}$ e $\lambda$ are scalar fields and
$\lambda^{i}$ is a divergenceless vector field. Now, under
infinitesimal coordinate transformations
\begin{equation}
\label{translcaiscosmol}
\tilde{\eta}=\eta+\xi^{0}, ~~~~~ \tilde{x}^{i}=x^{i}+\xi^{i},
\end{equation}
we can show that the following vector fields
\begin{equation}
\label{defvecpot} \varpi_i=S_i+F_i',~~~~ \sigma_i=\lambda_i-F_i,
\end{equation}
are invariant.

The action for Massive Gravity on a flat FRW background is then
given by (\ref{massgravaction}) with
(\ref{cosmopertmetricmassgrav}) and the Goldstone fields set to
their vacuum values (\ref{goldsvac}); the matter lagrangian
$\mathcal{L}_{matter}$ is assumed to be described by a perfect
fluid whose perturbations for the fluid four-velocity are
\begin{equation}
\nonumber
\delta u_i = a (\zeta_i + \partial_i \zeta),~~~
\delta u_0 = a\varphi.
\end{equation}

With these features, the Einstein equations for the tensor field
$\chi_{ij}$ are given by \cite{bebronne2007},
\begin{equation}\label{tensorpertmg}
{{\chi''}}_{ij}-\nabla^2 {{\chi}}_{ij} +2{\cal{H}}\chi_{ij}'+a^{2}m^{2}_{2}\chi_{ij}=0,
\end{equation}
whereas for the gauge-invariant vector fields defined by (\ref{defvecpot}) the Einstein equations read
\begin{equation} \label{cosmovecpertmg1}
\nabla^2 \varpi_i - 2 a^2 \rho_m M_{pl}^{-2} ( 1 + w)\zeta_i = 0,
~~~ \varpi_i' + 2 \mathcal{H} \varpi_i - a^2 m_2^2 \sigma_i = 0,
\end{equation}
\begin{equation}
\label{cosmovecpertmg3}
m_2^2 \nabla^2 \sigma_{i} = 0,
\end{equation}
where $\delta_\zeta = \zeta - \left( E^\prime + B \right)$, and $w$ is the parameter appearing the equation of state of the ordinary matter, $p = w\rho$ \cite{bebronne2007}.

Solving equations (\ref{cosmovecpertmg1}) -
(\ref{cosmovecpertmg3}) we conclude that the only relevant vector
field is $\varpi_i$, whose amplitude decays with $a^{-2}$
\cite{bebronne2007}, which is exactly the same behavior of vector
fields as derived in GR.

Therefore, the Fierz-Pauli modified model and Massive Gravity give
rise to the same results for the TT polarization modes of the
tensor perturbations as can be seen from equations
(\ref{tttensorpertmfp}) and (\ref{tensorpertmg}), whereas for
vector perturbations the situation changes drastically. In the
modified Fierz-Pauli model the vector modes of GW polarization
obey the same equation as the TT modes, (\ref{vectensorpertmfp}),
so that they really may contribute to the polarization of CMB as
we have shown in \cite{bessada2008} for the field $\chi_{ij}$;
however, in Massive Gravity, the vector perturbations behave
exactly as in GR, which means that they decay too fast after the
inflationary phase and do not leave any signature on CMB
polarization.

Therefore, the vector modes of the modified Fierz-Pauli model,
unlike the predictions of GR, Massive Gravity, and other modified
models of gravity, can leave signatures on the CMB. To see how
this can be achieved, let us estimate now a limit mass to be
detected by CMB polarization experiments, according to the
discussion in \cite{bessada2008}. Since the equations for the
tensor (\ref{tttensorpertmfp}), and vector
(\ref{vectensorpertmfp}) modes are identical, the dispersion
relations are also identical, given by
\begin{equation}\label{disprelation}
\omega^{2}=k^{2}+m^{2};
\end{equation}
now, since in GR only GW with frequencies $\nu$ within the range
$10^{-15}Hz$ to $10^{-18}Hz$ may leave a signature on CMB
polarization \cite{kamionkowski1998}, we see that these
frequencies correspond to comoving wavenumbers $k$ within the
range $10^{-25}cm^{-1}$ to $10^{-28}cm^{-1}$. We then use the
values of $k$ of GR, varying the frequencies in order to obtain
constant nonzero graviton masses through equation
(\ref{disprelation}). As a result, if the values of the graviton
mass $m$ lie within the range $10^{-66}$ - $10^{-62}g$, the
correspondent frequencies have values very close to the expected
in GR; in particular, there is a graviton mass limit,
$m=10^{-66}g\sim 10^{-29}cm^{-1}$, so that below this limit the
dynamical evolution of the massive modes is indistinguishable from
the tensor massless modes. Hence, since the vector modes in the
modified Fierz-Pauli model are governed by the same equation as
the tensor modes, then the graviton mass limit for the vector
modes should be exactly the same.

Along with the value discussed above for the vector massive modes,
there are other mass limits obtained so far in the literature by
using different tests. For instance, Goldhaber and Nieto
\cite{goldhaber1974} found a limit $m < 2.0 \times 10^{-62}g$
analyzing the motion of galaxies in clusters. Later on, Talmadge
\emph{et al.} \cite{talmadge1988} studied the variations of
Kepler's third law when compared with the orbits of Earth and
Mars, and found a limit $m < 7.68 \times 10^{-55}g$. Recently,
Finn and Sutton \cite{Finn2002} calculated the decay of the
orbital period of the binary pulsars PSR B1913+16 (Hulse and
Taylor pulsar) and PSR B1534+12 due to emission of massive
gravitons, and found $m < 1.4\times 10^{-52} g$.

\section{\label{sec:cmbpol}The Radiative Transfer Equation in the
 presence of Weak Gravitational Fields - an overview}

Once we have discussed the key features of the different theories
of gravitation with massive gravitons, we now turn our attention
to the polarization of CMB. To do so, we initially review the
theory of the radiative transfer in the presence of weak
gravitational fields, following closely the seminal paper by
Polnarev \cite{polnarev} (for a pedagogical introduction, see
\cite{kamionkowski2004}).

Let us consider a given beam of radiation characterized by its
\textit{Stokes parameters} $\{I_{l},I_{r},U\}$, where $I_{l}$ and
$I_{r}$ are the intensities of the radiation in the directions $l$
and $r$, respectively; $I=I_{l}+I_{r}$ is the total intensity of
the wave, and the parameter $Q$ is given by $Q=I_{l}-I_{r}$. These
functions are strictly related to the photon distribution
function, which can be cast in a symbolic vector of the form
\cite{chandrasekhar}
\begin{equation}\label{distfunct}
\hat{f}:=\left(
\begin{array}{c}
I_{l}\\
I_{r}\\
U\\
\end{array}
\right),
\end{equation}
where $U$ is the other parameter associated with linear
polarization. The components of $\hat{f}$ are functions of the
conformal time $\eta$, the comoving spatial coordinates
$\mathbf{r}$, and also of the photon angular distribution. Since
the photons are scattered by the free electrons prior to
recombination via Thomson scattering, their distribution function
will be shifted according to the equation \cite{polnarev}
\begin{equation}\label{boltz1}
\frac{\partial \hat{f}}{\partial \eta}+\frac{p^{i}
}{p}\frac{\partial \hat{f}}{\partial x^{i}}+\frac{\partial
\hat{f}}{\partial \nu}\frac{d \nu}{d \eta}=C[\hat{f}],
\end{equation}
where $\nu$ is the photon frequency, $\hat{\mathbf{p}}$ is the
photon momentum, $C[\hat{f}]$ is the scattering term given by
\begin{equation}\label{collterm}
C[\hat{f}]=-\sigma_{T}N_{e}a\left[\hat{f}-\frac{1}{4\pi}\int^{1}_{-1}d\mu'd\varphi
~ P\left(\mu,\varphi,\mu',\varphi'\right)\hat{f}\right],
\end{equation}
$\sigma_{T}$ is the Thomson scattering cross-section, $N_{e}$
is the number of free electrons in the unit comoving volume,
$\mu=\cos \theta$, and $P\left(\mu,\varphi,\mu',\varphi'\right)$
is the scattering matrix given by \cite{chandrasekhar}
\begin{eqnarray}\label{collterm1}
P\left(\mu,\varphi,\mu',\varphi'\right)&=&\tilde{P}\left[P^{0}\left(\mu,\mu'\right)+\sqrt{1-\mu^{2}}
\sqrt{1-\mu'^{2}}P^{1}\left(\mu,\varphi,\mu',\varphi'\right)\right. \\
\nonumber &+&\left.
P^{2}\left(\mu,\varphi,\mu',\varphi'\right)\right]
\end{eqnarray}
where
\begin{equation}\label{matscat1}
\tilde{P}=\left(
\begin{array}{cccc}
1 & 0 & 0 & 0 \\
0 & 1 & 0 & 0 \\
0 & 0 & 1 & 0 \\
0 & 0 & 0 & 2 \\
\end{array}
\right),~P^{0}=\frac{3}{4}\left(
\begin{array}{cccc}
2(1-\mu^{2})
(1-\mu'^{2})+\mu^{2}\mu'^{2} & \mu^{2} & 0 & 0 \\
\mu'^{2} & 1 & 0 & 0 \\
0 & 0 & 0 & 0 \\
0 & 0 & 0 & \mu \mu' \\
\end{array}
\right),
\end{equation}
\begin{equation}\label{matscat3}
P^{1}=\frac{3}{4}\left(
\begin{array}{cccc}
4\mu \mu' \cos \psi & 0 & -2\mu \sin \psi & 0 \\
0 & 0 & 0 & 0 \\
2\mu' \sin \psi & 0 & \cos \psi  & 0 \\
0 & 0 & 0 & \cos \psi \\
\end{array}
\right),
\end{equation}
\begin{equation}\label{matscat4}
P^{2}=\frac{3}{4}\left(
\begin{array}{cccc}
\mu^{2} \mu'^{2} \cos 2\psi & -\mu^{2}\cos 2\psi & -\mu^{2} \mu'
\sin
 2\psi & 0 \\
-\mu'^{2}\cos 2\psi & \cos 2\psi & \mu' \sin 2\psi & 0 \\
\mu \mu'^{2} \sin 2\psi & -\mu \sin 2\psi & \mu \mu' \cos 2\psi  & 0 \\
0 & 0 & 0 & 0 \\
\end{array}
\right),
\end{equation}
where we have defined $\psi:=\varphi-\varphi'$.

In equation (\ref{boltz1}) the third term on the left-hand side
shows the influence or signature of the GWs on the CMB
polarization, whereas the term on the right-hand side gives the
details of the Thomson scattering. Since the term (\ref{collterm})
depends upon the photon angular function through the angles
$\left(\mu,\varphi\right)$, we see that different GW imprints may
lead to different processes of polarization via Thomson
scattering. In the next two sections we address these issues in
detail.

\section{\label{sec:sachswolfe}The Sachs-Wolfe effect induced by Massive Gravitons}

As we have seen in Section \ref{sec:cosmopert}, the modified
Fierz-Pauli model and Massive Gravity are equivalent with respect
to the GW TT tensor modes, but only the vector modes of the first
model may give rise to relevant contributions to CMB polarization.
Hence, from this section on, we shall always consider
only the Fierz-Pauli model to address all the issues concerning GW
vector modes.

We have discussed in Section \ref{sec:cmbpol} that the CMB photons
are polarized due to the Thomson scattering with the free
electrons in the epoch of recombination. Prior to Thomson
scattering, the cosmological perturbations imprint a signature on
the photon angular pattern, the so-called \emph{Sachs-Wolfe} (SW)
effect, which can be understood as the shift of photon frequency
along the line of sight.

This effect can be computed through the geodesic equation for the
photon; then, using the metric perturbation (\ref{pertmetmodfp})
around a flat FRW space, and the constraint
$g_{\alpha\beta}p^{\alpha}p^{\beta}=0$, where $p^{\alpha}$ is the
photon four-momentum, the geodesic equation reads for the tensor
field $h_{ij}$,
\begin{eqnarray}\label{geodgen}
\frac{d\nu}{d\lambda}&=&-\nu \left[\mathcal{H}+\frac{1}{2}\frac{\partial
h_{ij}}{\partial \eta}p^{i}p^{j}\right]\frac{d \eta}{d\lambda},
\end{eqnarray}
where $\lambda$ is an affine parameter. The product $\varepsilon_{ij}p^ip^j$ appearing in (\ref{geodgen})
may be evaluated as follows: first, since the photon travels along
an arbitrary direction $\hat{\mathbf{p}}$, let us construct a
reference frame around the GW propagation vector
$\hat{\mathbf{k}}$ such that the GW polarization modes assume the
simplified form given by (\ref{polariztensor}). To this end we
introduce the \textit{polarization vectors}
$\{{\hat{\varepsilon}}^r_{(1)},{\hat{\varepsilon}}^r_{(2)}\}$,
defined by
\begin{equation}\label{polvectors}
\varepsilon^r_{ij}=\varepsilon^r_{(1)i}\varepsilon^r_{(1)j}-\varepsilon^r_{(2)i}\varepsilon^r_{(2)j},
\end{equation}
where $\varepsilon^r_{ij}$ is given by (\ref{polariztensor}) for
each polarization component $r$, satisfying
\begin{equation}\label{gwbasisprop}
{\hat{\varepsilon}}^r_{(1)}\cdot
{\hat{\varepsilon}}^r_{(2)}={\hat{\varepsilon}}^r_{(1)}\cdot
\hat{\mathbf{k}}={\hat{\varepsilon}}^r_{(2)}\cdot
\hat{\mathbf{k}}=0.
\end{equation}
The trihedron
\begin{equation}\label{gwbasis}
\{{\hat{\varepsilon}}^r_{(1)},{\hat{\varepsilon}}^r_{(2)},\hat{\mathbf{k}}\}
\end{equation}
is then a reference frame around the GW direction
$\hat{\mathbf{k}}$ in which the polarization tensor preserves the
simple form (\ref{polariztensor}).

Second, let us express the photon momentum $\mathbf{p}$ in terms
of spherical coordinates around (\ref{gwbasis}), so that the
following relations hold,
\begin{equation}\label{photonmom}
\hat{\mathbf{k}}\cdot \hat{\mathbf{p}}=\cos \theta,~~~
{\hat{\varepsilon}}^r_{(1)}\cdot \hat{\mathbf{p}}=\sin\theta \cos
\varphi,~~~ {\hat{\varepsilon}}^r_{(2)}\cdot
\hat{\mathbf{p}}=\sin\theta \sin\varphi,
\end{equation}
therefore, using (\ref{polariztensor}) and
(\ref{photonmom}), it follows that
\begin{equation}\label{prodepsprepsi3}
\varepsilon^2_{ij}p^ip^j=\mu\sqrt{1-\mu^{2}} \cos\varphi \propto
Y_{2,+1}\left(\mu,\varphi\right),
\end{equation}
\begin{equation}\label{prodepspimpsi3}
\varepsilon^3_{ij}p^ip^j=\mu\sqrt{1-\mu^{2}} \sin\varphi\propto
Y_{2,-1}\left(\mu,\varphi\right),
\end{equation}
\begin{equation}\label{prodepsprepsi4}
\varepsilon^4_{ij}p^ip^j=\left(1-\mu^{2}\right) \cos
2\varphi\propto Y_{2,+2}\left(\mu,\varphi\right),
\end{equation}
\begin{equation}\label{prodepspimpsi4}
\varepsilon^5_{ij}p^ip^j=\left(1-\mu^{2}\right) \sin
2\varphi\propto Y_{2,-2}\left(\mu,\varphi\right);
\end{equation}
where $Y_{lm}\left(\mu,\varphi\right)$ are the usual spherical
harmonics. The results (\ref{prodepsprepsi3}) -
(\ref{prodepspimpsi4}) show us that the GW imprint upon the photon
angular distribution is in the form of a quadrupole, with $m=\pm
2$ for the $\Psi_4$  modes ($r=4,5$, which coincides with GR), and
with $m=\pm1$ for the $\Psi_3$ modes ($r=2,3$).

Hence, from
relations (\ref{cosmofourierexph}), (\ref{cosmofourierexph1}), (\ref{geodgen}) and
(\ref{prodepsprepsi3})-(\ref{prodepspimpsi4}) we get the following
geodesic equations:

\textit{i}) $\Psi_{3}$ :
\begin{equation}\label{geodpsi3}
\frac{1}{\nu_0}\frac{d\nu_0}{d\eta}\propto-\frac{\partial h^{2,3}}{\partial
\eta}
Y_{2,\pm1}\left(\mu,\varphi\right),
\end{equation}

\textit{ii}) $\Psi_{4}$ :
\begin{equation}\label{geodpsi4}
\frac{1}{\nu_0}\frac{d\nu_0}{d\eta}\propto-\frac{1}{2}\frac{\partial h^{4,5}}{\partial
\eta}
Y_{2,\pm 2}\left(\mu,\varphi\right)
\end{equation}
where $\nu_0=\nu a(\eta)$.

Then, massive gravitons with the $\Psi_{4}$ polarization modes give rise to the usual tensor SW effect in GR, whereas the $\Psi_{3}$ modes do not yield its well-known vector version in GR (see, for example, the discussion in \cite{giovannini2005}). This happens because $\Psi_{3}$ modes arise as GW longitudinal states of polarization, and not as a massive vector fields as in GR or Massive Gravity. Anyway, they will leave a different signature on the CMB polarization by means of Thomson scattering, as we discuss in the next section.

\section{\label{sec:thomscat}The Basis for Thomson Scattering}

We now turn to the derivation of the Thomson scattering term
(\ref{collterm}) for massive gravitons. Let us consider first that
the incident radiation prior to be Thomson scattered is
unpolarized, with the angular pattern dictated by the SW effect
(\ref{geodpsi3}) and (\ref{geodpsi4}). In this case, the Stokes
vectors for the incident radiation are given by

\textit{i}) $\Psi_{3}$ :
\begin{equation}\label{apsi3}
{\hat{a}}^{2}=\frac{1}{2}\mu \sqrt{1-\mu^{2}} \cos \varphi ~\mathbf{\hat{u}},~~~
{\hat{a}}^{3}=\frac{1}{2}\mu \sqrt{1-\mu^{2}} \sin \varphi ~\mathbf{\hat{u}},
\end{equation}

\textit{ii}) $\Psi_{4}$ :
\begin{equation}\label{apsi4}
{\hat{a}}^{4}=\frac{1}{2}\left(1-\mu^{2}\right) \cos 2\varphi ~
\mathbf{\hat{u}},~~~ {\hat{a}}^{5}=\frac{1}{2}\left(1-\mu^{2}\right) \sin
2\varphi ~\mathbf{\hat{u}},
\end{equation}

where we have defined
\begin{equation}\label{defvecu}
\mathbf{\hat{u}}=
\left(
\begin{array}{c}
1\\
1\\
0\\
\end{array}
\right).
\end{equation}

Now, defining the operator $\hat{P}$ as
\begin{equation}\label{poper}
\hat{P}\hat{\xi}\left(\mu,\varphi\right)=\frac{1}{4\pi}\int^{1}_{-1}d\mu'd\varphi'
~
P\left(\mu,\varphi,\mu',\varphi'\right)\hat{\xi}\left(\mu',\varphi'\right),
\end{equation}
where $P$ is the scattering matrix (\ref{collterm1}), it is
 straightforward to see, for $\hat{\xi}={\hat{a}}^{r}$ ($r=2,3,4,5$), that
\begin{equation}\label{defbasis1}
\hat{P}{\hat{a}}^{r}=p {\hat{a}}^{r}+q {\hat{b}}^{r},
\end{equation}
where $p$ and $q$ are constants, and ${\hat{b}}^{r}$ is a
basis vector such that
\begin{equation}\label{defbasis2}
\hat{P}{\hat{b}}^{r}=p '{\hat{a}}^{r}+q' {\hat{b}}^{r},
\end{equation}
where $p'$ and $q'$ are constants as well. From (\ref{apsi3}) - (\ref{defbasis2}) we readily see that,

\textit{i}) $\Psi_{3}$ :
\begin{equation}\label{bpsi3}
{\hat{b}}^{2}=\frac{1}{2}\sqrt{1-\mu^{2}} \left(
\begin{array}{r}
\mu \cos \varphi \\
-\mu  \cos \varphi \\
2\sin \varphi \\
\end{array}
\right),~~~
{\hat{b}}^{3}=\frac{1}{2} \sqrt{1-\mu^{2}} \left(
\begin{array}{r}
\mu\sin \varphi \\
-\mu \sin \varphi \\
-2\cos \varphi \\
\end{array}
\right),
\end{equation}

\textit{ii}) $\Psi_{4}$ :
\begin{equation}\label{bpsi4}
{\hat{b}}^{4}=\frac{1}{2}\left[
\begin{array}{c}
\left(1+\mu^{2}\right) \cos 2\varphi \\
-\left(1+\mu^{2}\right) \cos 2\varphi \\
4\mu  \sin 2\varphi \\
\end{array}
\right],~~~ {\hat{b}}^{5}=\frac{1}{2}\left[
\begin{array}{c}
\left(1+\mu^{2}\right) \sin 2\varphi \\
-\left(1+\mu^{2}\right) \sin 2\varphi \\
-4\mu \cos 2\varphi \\
\end{array}
\right].
\end{equation}

The basis vectors given by (\ref{apsi3}) and (\ref{bpsi3}),
(\ref{apsi4}) and (\ref{bpsi4}) allows us to factor out the
angular dependence of the photon distribution vectors, and they
constitute the first-order contribution to $\hat{f}$. The
remaining contribution for $\hat{f}$ comes from the $h=0$
solution, that is, in the absence of GWs, in which the photon
distribution vector is given by
\begin{equation}\label{f0}
\hat{f}:={\hat{f}}_{0}=f_{0}(\nu_{0})~\mathbf{\hat{u}}.
\end{equation}
Now, the linearized photon distribution vector can be written as
\begin{equation}\label{phdistvec}
{\hat{f}}^{r}\sim
{{\hat{f}}_{0}}^{r}+e^{-i\vec{k}\cdot\vec{r}}\left[\alpha^{r}\left(\eta,\mu,\nu_{0}
\right){\hat{a}}^{r}+\beta^{r}
 \left(\eta,\mu,\nu_{0} \right){\hat{b}}^{r}\right],
\end{equation}
where $\alpha^{r}\left(\eta,\mu,\nu_{0} \right)$ and $\beta^{r}\left(\eta,\mu,\nu_{0} \right)$ are functions to be determined as solutions of the Boltzmann's equations (\ref{boltz1}).

\section{\label{sec:boltzeq}The Complete Boltzmann Equations}

Now, once we have obtained the form of the photon distribution
vector (\ref{phdistvec}) for the TT and longitudinal GW mode, we are able to write down the full Boltzmann equations (\ref{boltz1}).
They are given by

\textit{i}) $\Psi_{3}$ :
\begin{equation}\label{boltzpsi31}
{\chi^{r}}' +(q-ik\mu)\chi^{r} =H^{r},
\end{equation}
\begin{eqnarray}\label{boltzpsi32}
{\beta^{r}}'&+&(q-ik\mu)\beta^{r} =-\frac{3}{8}q
\int^{1}_{-1} d\mu'
 \left[-\chi^{r} \mu'^{2}(1-\mu'^{2})\right. \nonumber \\&+& \left. \beta^{r}\left(-1-\mu'^{2}+2\mu'^{4}\right) \right],
\end{eqnarray}
for $r=2,3$, and

\textit{ii}) $\Psi_{4}$ :
\begin{equation}\label{boltzpsi41}
{\xi^{r}}'+(q-ik\mu)\xi^{r} =H^{r},
\end{equation}
\begin{eqnarray}\label{boltzpsi42}
{\beta^{r}}'&+&(q-ik\mu)\beta^{r} =\frac{3}{16}q
\int^{1}_{-1} d\mu'
\left[\beta^{r}(1+\mu'^{2})^{2}-\frac{1}{2}\xi^{r}
 (1-\mu'^{2})^{2} \right],
\end{eqnarray}
for $r=4,5$. The functions $\xi$ and $\chi$ are defined as
\begin{equation}\label{defxi}
\xi^{r}=\alpha^{r}+\beta^{r},
\end{equation}
\begin{equation}\label{defchi}
\chi^{r}=\alpha^{r}-\beta^{r},
\end{equation}
and the scattering rate $q$ is defined as $q=\sigma_{T}N_{e}a$,
and finally
\begin{equation}\label{defH}
H^{r}=\frac{1}{2}\frac{\partial h^{r}(\eta)}{\partial
 \eta}.
\end{equation}

The equations (\ref{boltzpsi41}) and (\ref{boltzpsi42}) for
massive gravitons with $\Psi_4$ mode are identical to the
correspondent ones in GR; the difference here lies on the
Boltzmann equations for the $\Psi_3$ modes, (\ref{boltzpsi31}) and
(\ref{boltzpsi32}), which do not appear in GR. Since the vector
and tensor modes satisfy the same dynamical equation, and the
mathematical form of the equations (\ref{boltzpsi31}) and
(\ref{boltzpsi32}) is very different from (\ref{boltzpsi41}) and
(\ref{boltzpsi42}), it is clear that the vector polarization modes
of massive gravitons leave a characteristic signature
distinguishable from the tensor one, which could, in principle, be
probed by measurements on the CMB $E$ and $B$-modes. Since the
experiments in the Planck satellite will improve the WMAP5 results
for the $E$-mode, we may expect that such future measurements
might decide whether nontrivial GW signatures - as we showed here
through equations (\ref{boltzpsi31}) and (\ref{boltzpsi32}) for
$\Psi_3$-modes - appear or not in the CMB polarization spectrum.
In this case, we conclude that CMB polarization measurements may
be decisive to test alternative theories of gravitation - in
particular, the massive model as we discussed here.

\section{\label{conc}Conclusions}

We have analyzed in this work three theories of gravitation with
massive gravitons, and we have shown that the modified Fierz-Pauli
model coincides with Massive Gravity for the TT tensor field in
linear perturbation theory around Minkowski and flat FRW
backgrounds. We have shown that such tensor perturbations
associated with massive gravitons give rise to the usual tensor SW
effect, lead to the same Boltzmann equations for these modes in
GR.

We have deduced the dynamical equations for the GW vector
longitudinal polarization modes ($\Psi_3$-modes) in the modified
Fierz-Pauli model and shown that they do not give the same results
of Massive Gravity, in which vector perturbations behave like in
GR; instead, they give rise, in a cosmological scenario, to a
nontrivial SW effect which leaves a vector signature of the
quadrupolar form $Y_{2,\pm 1}(\mu,\varphi)$ on the CMB
polarization. Also, such massive vector modes possess a mass limit
of $m\sim 10^{-29}cm^{-1}$ as the tensor modes, so that below this
limit the Fierz-Pauli model is absolutely indistinguishable from
GR in terms of the dynamical evolution.

Analyzing the Einstein equations for such $\Psi_3$-modes we
concluded that these vector signatures could be present at
recombination epoch, unlike the vector perturbations in GR and
Massive Gravity, which would decay too fast and would not leave
any signature on CMB polarization. Therefore, we calculated the
new basis for the Thomson scattering for such $\Psi_3$-modes, and
then deduced the appropriate equation for the radiative transport.
Based upon these results we have shown qualitatively that
$\Psi_{3}$-mode vector signatures could clearly be distinguished
on the CMB polarization from the usual $\Psi_4$ tensor modes if
the former do exist; hence, we could look for such signatures in
the $E$-mode performed by Planck satellite.

In this sense we argued that Planck polarization measurements
could be decisive to test alternative theories of gravitation.

\ack
DB and ODM thanks Odylio Aguiar, Armando Bernui, Thyrso Villela
and Carlos Alexandre Wuensche for very helpful discussions. The
authors also thank Professor Jos\'e A. de Freitas Pacheco for very
important discussions and for a critical reading of the
manuscript. DB also thanks Professor Mark Kamionkowski for
clarifying some important points concerning CMB polarization. DB
was financially supported
 by CAPES, and ODM is partially supported by CNPq (grant 305456/2006-7).


\newpage

\section*{References}

\begin{thebibliography}{99}

\bibitem{lightman}
D. M. Eardley, D. M. Lee and A. P. Lightman. Phys. Rev. D
\textbf{8}, 3308 (1973)

\bibitem{miranda}
C. Stellati, O. D. Miranda and R. M. Marinho. \emph{In
preparation}.

\bibitem{Finn2002}
P. J. Sutton and L. S. Finn. Phys. Rev. D \textbf{65}, 044022
(2002)

\bibitem{gruzinov1}
G. Gabadadze and A. Gruzinov. Phys. Rev. D \textbf{72} 12, 124007 (2005)

\bibitem{visser98}
M. Visser. Gen. Rel. and Grav. \textbf{30}, 1717 (1998)

\bibitem{rubakov2004}
V. Rubakov.: Lorentz-violating graviton masses: getting around
ghosts, low strong coupling scale and VDVZ discontinuity.
\emph{hep-th/0407104} (2004)

\bibitem{dubovsky2004}
S. L. Dubovsky. JHEP \textbf{10}, 76 (2004)

\bibitem{fierz1939}
M. Fierz and W. Pauli. Proc.
Roy. Soc. Lond. \textbf{A}, 173 (1939)

\bibitem{boulware}
D.G. Boulware and S. Deser. Phys. Rev. D \textbf{6}, 3368 (1972)

\bibitem{rosen1973}
N. Rosen. Ann. of Phys. \textbf{84}, 455 (1973)

\bibitem{wayne2004}
W. L. S. de Paula, O. D. Miranda and R. M. Marinho. Class. Quantum
Grav. \textbf{21}, 4595 (2004)

\bibitem{Veltman1970}
H. van Dam and M. Veltman. Nucl. Phys. \textbf{B22}, 397 (1970)

\bibitem{zakharov}
V.I. Zakharov. JETP Lett. \textbf{12}, 312 (1970)

\bibitem{ligo}
B.F. Schutz. Class. Quantum Grav. \textbf{16}, A131, (1999)

\bibitem{bmode}
M. Tucci \emph{et. al}. MNRAS \textbf{360},  935 (2005)

\bibitem{penrose}
E. Newman and R. Penrose. J. Math. Phys \textbf{3}, 566 (1962)

\bibitem{will1}
C. M. Will.: The Confrontation between General Relativity and
Experiment. Living Reviews in Relativity \textbf{3}, 9
http://www.livingreviews.org/lrr-2006-3 (2006)

\bibitem{dubovsky2005a} S. L. Dubovsky, P. G. Tinyakov and I. I. Tkachev. Phys. Rev. D \textbf{72}, 084011 (2005)

\bibitem{brandenberger1992}
V. F. Mukhanov,  H. A. Feldman and R. H. Brandenberger.: Theory of cosmological perturbations. Phys. Rep. \textbf{215}, 203-333 (1992)

\bibitem{bebronne2007} M. V. Bebronne and P. G. Tinyakov. Phys. Rev. D \textbf{76}, 084011 (2007)

\bibitem{bessada2008}
D. Bessada and O. D. Miranda.: Polarization of CMB induced by tensor modes of primordial GWs in Massive Gravity \emph{Submitted to Class. Quantum Grav.} (2008)

\bibitem{kamionkowski1998} R.R. Caldwell, M. Kamionkowski and L.
Wadley, \emph{Phys. Rev D} \textbf{59}, 027101, (1998).

\bibitem{goldhaber1974} A. S. Goldhaber and M. M. Nieto.: M. M. Mass of the graviton.
Phys. Rev. D \textbf{9}, 1119, (1974).

\bibitem{talmadge1988} C. Talmadge, J. P. Berthias, R. W. Hellings
and E. M. Standish.: Model-independent constraints on possible
modifications of newtonian gravity. Phys. Rev. Lett. \textbf{61},
n. 10, (1988).

\bibitem{polnarev}
A. G. Polnarev. Sov. Astron. \textbf{29}(6), 307 (1985)

\bibitem{kamionkowski2004} P. Cabella and M. Kamionkowski.: Theory of Cosmic Microwave Background Polarization \emph{arXiv: astro-ph/0403392} (2004)

\bibitem{chandrasekhar}
S. Chandrasekhar.: Radiative Transfer, \textit{Dover Ed.} (1960)

\bibitem{giovannini2005}
M. Giovannini.: Theoretical Tools for CMB Physics. IJMP D \textbf{14}, 363 (2005)


\end{thebibliography}
\end{document}